# Optical linear-nonlinear and dispersion parameters of thermally evaporated SnS thin films as absorber material for solar cells

Vinita[1], P. Arun[2], Chandra Kumar[3], R. Rai[1$], B.K. Singh[1*]

[1]*Department of Physics, Banaras Hindu University (BHU), Varanasi-221005, India
[2]Department of Electronics, SGTB Khalsa College, University of Delhi, Delhi 110007, India
[3]Departamento de Física, Facultad de Ciencias, Universidad Católica del Norte, Avenida Angamos 0610, Casilla 1280, Antofagasta, Chile
[$]Present address: INFN Sezione di Trieste, Via A. Valerio, 2 34127 Trieste, Italy

## Abstract:

In this manuscript, we report the results of optical properties of SnS thin films, deposited on FTO coated glass substrates at room temperature by thermal evaporation technique. In addition, the effect of film thickness on the optical behavior of FTO/SnS is analyzed and obtained results are compared with data of SnS films grown on glass and ITO substrates. Our study indicates that the properties of SnS film are independent of the substrate material. Further, the influence of the film thickness on the other optical parameters including, linear and third order nonlinear optical constants and dispersion parameters have also been investigated using the transmission, reflection, and absorption spectra. It is found that the optical band gap decreases from 2.07 to 1.30 eV with increase in SnS film thickness, whereas the refractive index increases with increasing thickness. Additionally, the oscillator energy, and the dispersion energy are estimated using Wemple–DiDomenico approach. The dispersion energies are in the range of 7.20–4.59 eV, while the oscillator energies of the thin films are in the range of 5.49–2.24 eV. Moreover, the nonlinear refractive index, and optical susceptibility are calculated by using the empirical relation of Tichy and Ticha. The volume of data suggests optical properties of SnS thin films are strongly dependent on film thickness.

# 1. Introduction

Tin sulfide (SnS) is an interesting IV-VI semiconductor with an orthorhombic structure, which has received significant attention over the few past decades due to its desirable properties, such as high optical absorption coefficient, narrow band gap and high native free-carrier concentration [1]. Such properties of SnS makes it an attractive candidate for various applications such as, photo-catalysis [2], gas sensors [3], photo-electrochemical [4], photodetector [5], solar cells [6][7], Schottky diodes [8]. In addition, SnS is widely used as an alternative solar absorber layer to conventional thin film absorbers such as Copper Gallium Selenide (CGS), Copper Zinc Tin Sulfide (CZTS), Copper Indium Gallium Selenide (CIGS), Cadmium Telluride (CdTe), and Copper Indium Gallium Selenide (CIGS) due to its several advantages, properties including biological non-toxicity [9] [10].

Since the deposition technique of thin films strongly influence the physical and chemical properties, researchers have studied SnS films using various deposition techniques, such as chemical bath deposition [5], sol gel spin coating technique [11], microwave hydrothermal [12], spray pyrolysis [13], atomic layer deposition [14], electron beam evaporation [15], DC/RF magnetron sputtering [16], pulsed laser deposition [17] and thermal evaporation [18]. T. Garmim et. al. [11] reported the sol gel spin coating synthesis of SnS thin films with different annealing temperatures and found that as annealing temperature increases, the band gap decreases and refractive index increases. Hyeongsu Choi et. al. [19] reported the atomic layer deposition of SnS thin films with different annealing temperatures and investigated the role of seed layer on the growth of SnS thin film. Among the listed techniques, the thermal evaporation technique is the most suitable physical deposition technique due to its large area coverage, comparatively high stability, high deposition rate, high reproducibility and control growth i.e. morphology, size and structure [20].

SnS elements consist of Sn and S, both are abundant and less toxic in nature. SnS films with Sn vacancies is a p-type semiconductor with a direct energy bandgap of 1.35 eV, which is near the optimum energy bandgap (1.5 eV) of solar cells, which makes it an interesting and desirable candidate for potential application as a solar absorber layer in thin film based photovoltaic devices [20]. However, most of these works only deal with the linear properties (optical) of SnS thin films and so far, the nonlinear optical parameters are not addressed well. For a better understanding of the parameters which control the optical band-gap, study of other linear and nonlinear optical parameters should be undertaken. With this motivation, the present study is aimed to investigate the effect of film thickness on the linear-nonlinear optical constants and dispersion parameters of SnS thin films deposited on FTO substrates by thermal evaporation technique. The optical measurements like transmission, reflection and absorption of various SnS thin films have been used to estimate the band gap, Urbach tail, complex refractive index, the dispersion energy parameters (dispersion energy and oscillator energy), lattice energy, carrier concentration, plasma frequency, third order of nonlinear optical susceptibility and nonlinear refractive index. Such optical parameters may provide us more clear insight to understand the influence of thickness on solar device performances.

## 2. Experimental details

SnS thin films were deposited using thermal evaporation technique onto FTO coated glass substrates. The FTO coated glass substrates have been cleaned in ethanol, and deionized water for 15 minutes. SnS powder of 99.5% (Alfa Aesar) purity was used as the source material, which has been kept at the 13 cm away from the substrate holder in the vacuum chamber. The evaporation current was 49.4 A, and a bias voltage was around 60 V. The deposition of samples has been carried out at room temperature under a high vacuum of $10^{-6}$ mbar. The rate of deposition of the films was kept at ~3.5 A°/s and the films thickness (from 100 to 600 nm) was controlled with the help of a thickness monitor (Sycon Model: STM-100/ MF).

The optical properties of the SnS thin films were observed by UV-Visible spectroscopy (Model: Jasco V-770 spectrophotometer). Surface morphology was determined by scanning Electron Microscope (Model: EVO 18 Research). The structural properties of SnS thin films were studied by X-ray diffraction (Model: Bruker D8 Advance ECO) with the radiation source of Cu K-alpha (0.154 nm). The elemental composition of Sn and S was studied by energy dispersive X-rays (Model: EDX) (Octane Pro).

## 3. Results and discussion

### 3.1 Microstructural Analysis of SnS Thin Films:

Fig. 1(a) shows the XRD patterns of SnS thin films with thickness of 600 nm deposited onto FTO substrate by thermal evaporation technique. All SnS peaks in the pattern are closely matched to those of orthorhombic SnS corresponding to JCPDS card number 39-0354. The peaks observed for the deposited SnS thin films were at $2\theta$, 27.43°, 30.89°, 31.82°, 39.15°, 45.84°, 51.82°, 53.28°, 64.50° and 78.63°. These peaks can be assigned to (021), (101), (040), (041), (002), (151), (160), (251) and (341) planes, respectively. Additional peaks are that appeared at $2\theta$, 26.80°, 34.07°, 38.08°, 61.84°, 65.82° and 80.85°, respectively correspond to FTO substrate [JCPDS card number 39-0354].

The surface morphology of the SnS thin films for sample with thickness 600 nm is shown in Fig.1(b). The deposited SnS thin films clearly exhibit flake like grain morphology, that is homogeneous and with a surface that is uniform and crack free. The average particle size of the SnS thin films were calculated through ImageJ software. The EDS spectra of SnS thin films for samples with thickness of 600 nm is shown in Fig.1(c). From the recorded EDS spectra, various elements such as Sn, S, O and C were detected. Fig.1(d) show the 2D atomic force microscope (AFM) topographical images of SnS thin films with thickness of 600 nm onto FTO substrate and with a scan size of $5\mu m \times 5\mu m$. The parameters like average roughness ($R_a$) and root mean squire roughness ($R_{rms}$) are also evaluated using Gwyddion software. The values of parameters $R_a$ and $R_{rms}$ are 24.67 nm and 29.80 nm respectively.

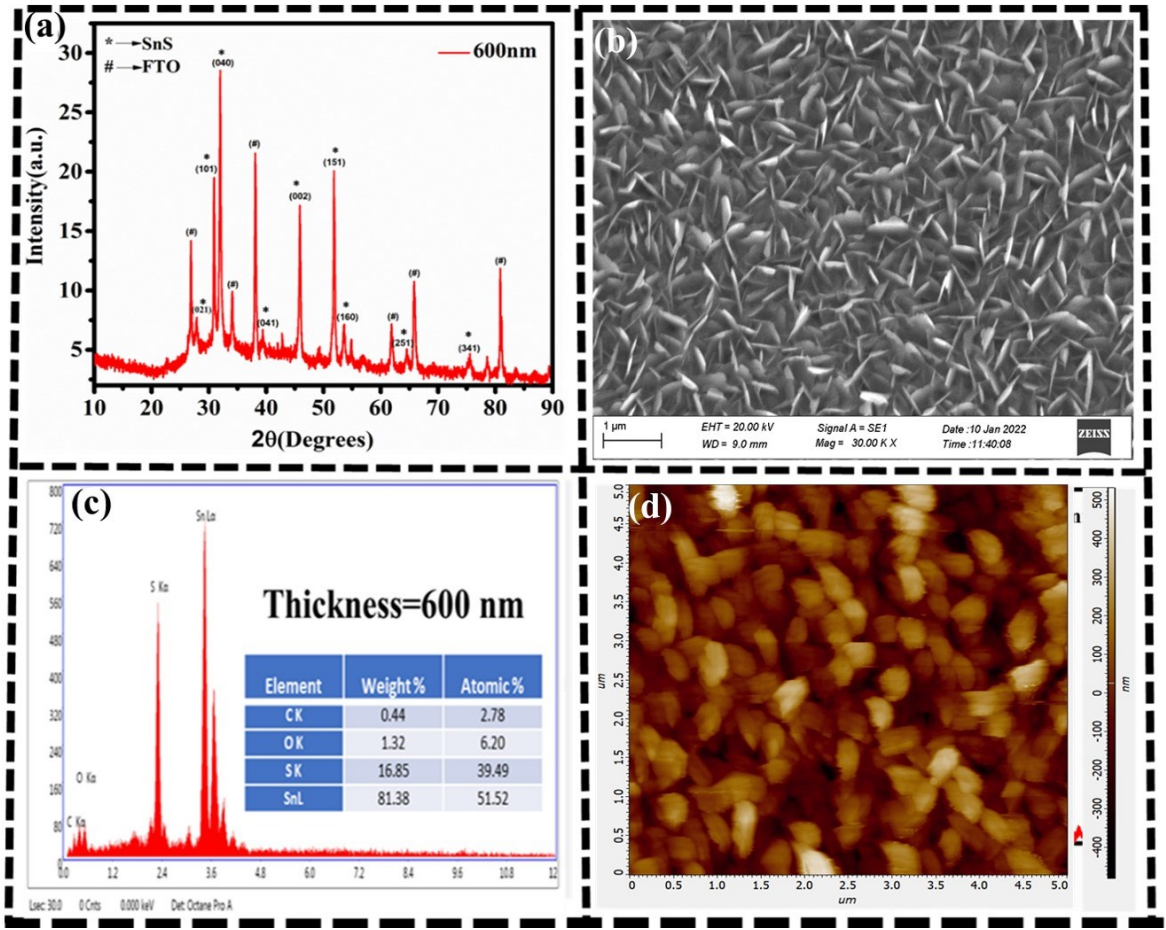

Fig.1:(a) XRD (b) SEM (c) EDS and (d) AFM of SnS thin film with thickness of 600 nm.

XRD, EDS and SEM analysis confirm the formation of uniform and high quality SnS films on FTO substrates at room temperature by thermal evaporation technique. In the subsequent section, the optical properties of these samples are discussed in detail.

### 3.2 Optical Analysis of SnS Thin Films:

### 3.2.1 Transmission and Absorption Analysis of SnS Thin Films:

The transmission spectra of SnS thin films with different thickness deposited onto FTO substrates in the range of 100–600 nm is shown in Fig. 2(a). The transmittance spectra show excellent transparency for wavelengths greater than 900-1000 nm. However, the strong absorption of SnS thin films in the visible region are acceptable for various opto-electronics applications such as solar power system, anti-reflective coating etc. The fringes observed is due to the interference between the waves reflected by the SnS surface and FTO substrate-film interfaces [21]. Here as film thickness increases, the transmittance decreases due to the increase in reflectance and absorbance of SnS thin films. A very weak transmittance at around 500 to 850 nm, which is close to 0% was also observed for all samples. The region of strong absorbance for all films was observed in this region due to the excitation and the migration of the electrons from the valence band to the conduction band [22]. The absorption spectra of the SnS films deposited at thickness 100 to 600 nm in the range of 550–2200 nm is shown in Fig. 2(b). Analysis indicates that the absorbance of SnS thin films decreases sharply at a critical wavelength (around 850 to 1100 nm for all samples), and then becomes relatively constant (wavelength regions > energy bandgap). Moreover, optical absorption edge shifts towards higher wavelength, indicate red shift in band-gap with increasing thickness. Such a shift at the absorption edge are reported for various thin films like SnS [23-24], Cu:ZnS [25]. Tauc method was used to determine the band-gap of the films, the obtained results are discuss in the next section.

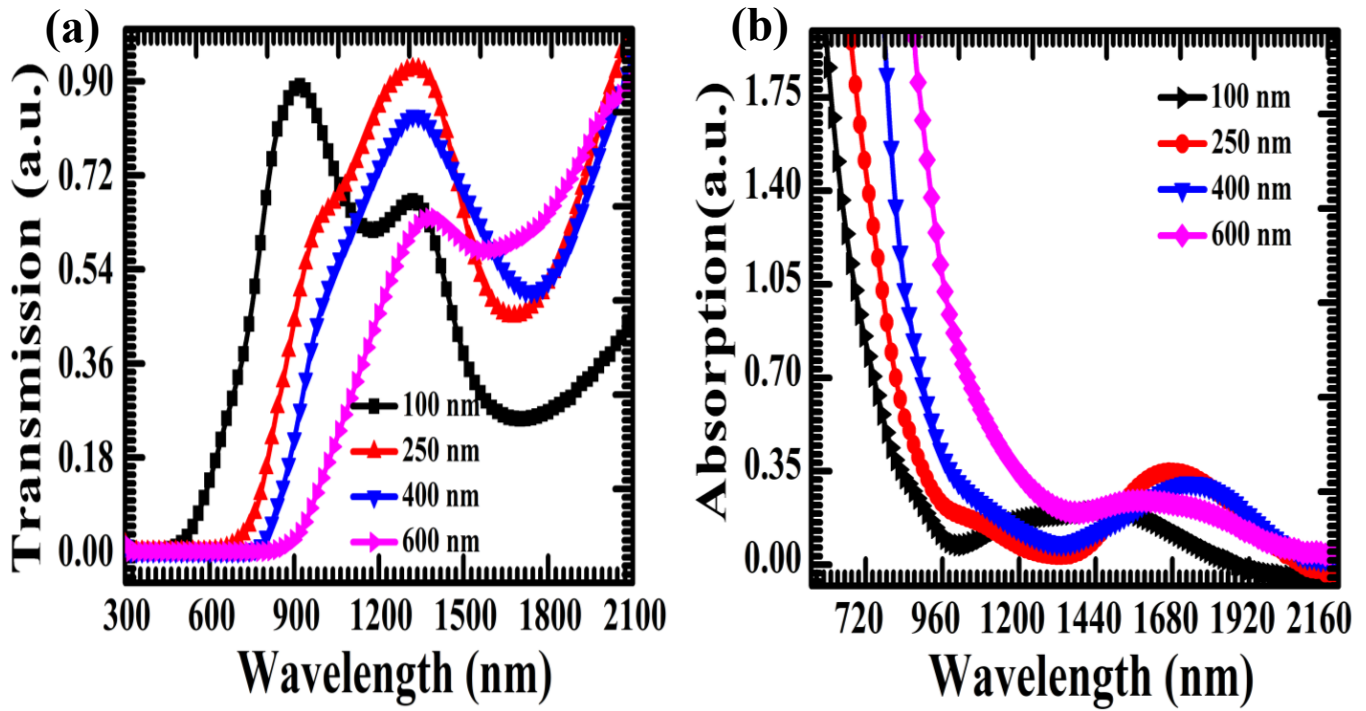

Fig.2(a) Transmission spectra, (b) Absorption spectra of SnS thin film with thickness of 100 to 600 nm.

### 3.2.2 Optical Band Gap, Urbach Tail and Effective Mass of SnS Thin Films

The optical band gap ($E_g$) of SnS thin films with different thicknesses are estimated using Tauc's plot [26]:

$$\alpha h\nu = B(h\nu - E_g)^n \tag{1}$$

Where, B is the constant, α is the absorption coefficient, hυ is the energy of incident photon, n is power factor. The value of n depends on transitions, for direct allowed, direct forbidden, indirect allowed and indirect forbidden transition n will be 1/2, 3/2, 2 and 3 [27]. The plot between $(\alpha h\nu)^2$ and the photon energy (hυ) for the SnS thin films deposited onto FTO substrate is shown in Fig. 3. The estimated optical band gap lies in range of 2.07 to 1.3 eV for all samples are summarized in Table 1. It can be perceived from results that the optical band gap decreases with increasing film thickness. Similar band gap range for SnS thin films are reported in literature [28-29].

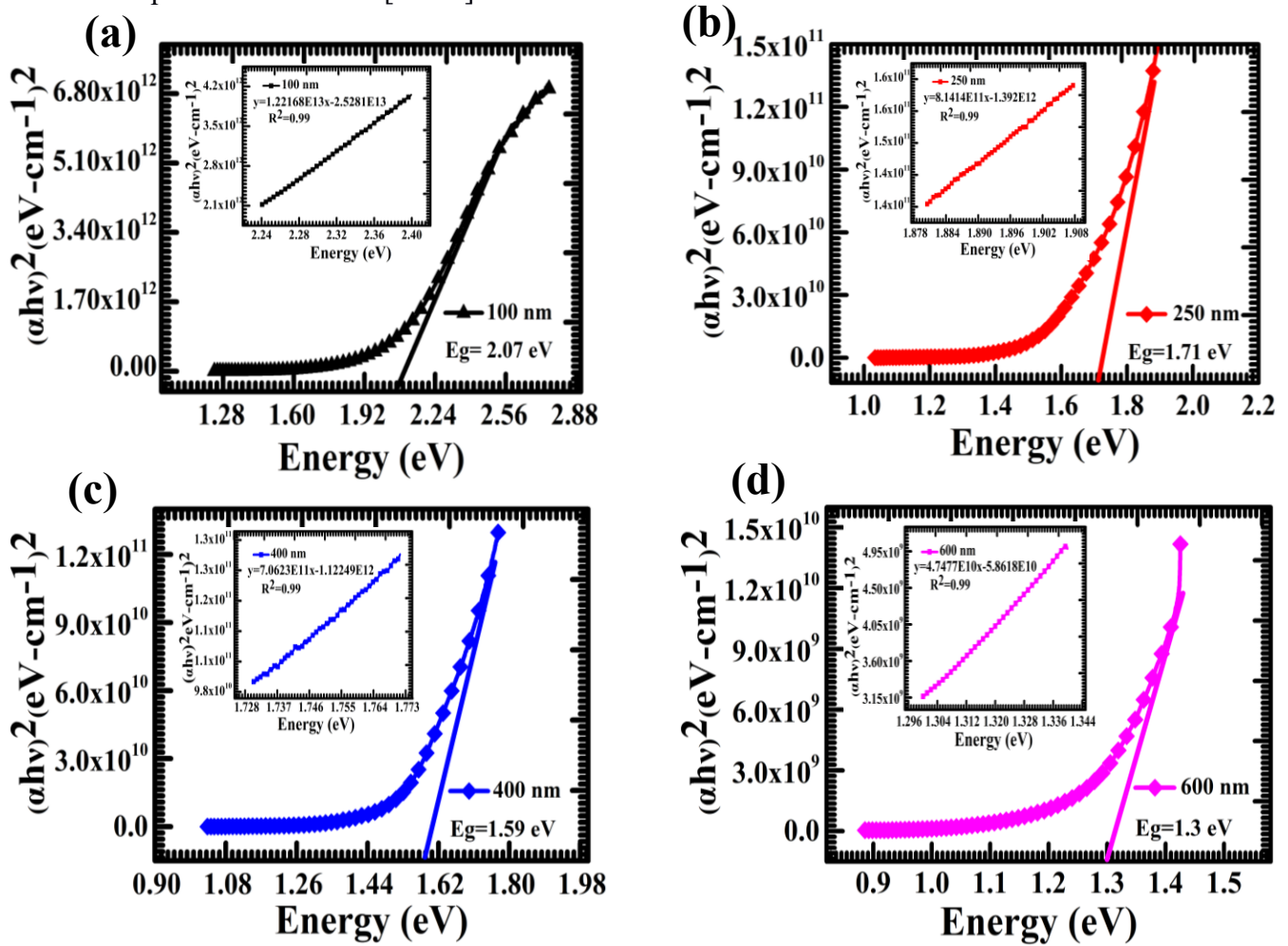

Fig.3 Optical band gap of SnS thin film with thickness of 100 to 600 nm.

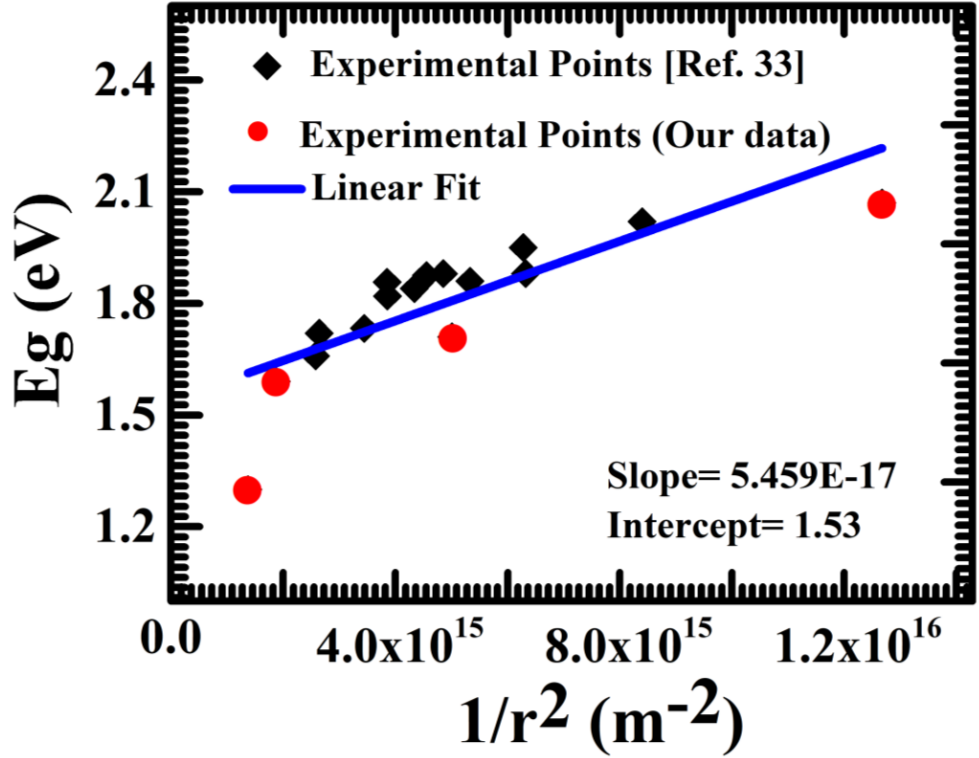

Fig.4: Variation of band gap with grain size with different thickness of SnS thin film.

The red-shift can be explained due to quantum confinement. Brus (1984) reported a theoretical model for explaining the variation of a bandgap with grain size. The result of his model is mathematically expressed as [30,31,32]:

$$E_g^* = E_g^{bulk} + \frac{\hbar^2 \pi^2}{2r^2}\left(\frac{1}{m_h^*}\right) \qquad (2)$$

where 'r' is the radius, $m_h^*$ is effective mass of hole (Since our samples are p-type, so the effective mass of electron ($m_e^*$) can be neglected). The first term in above equation gives the band-gap of a thick sample that can be considered as bulk while the second term $\frac{\hbar^2 \pi^2}{2}\left(\frac{1}{m_h^*}\right)$ represents confinement of charge carriers with the grain boundaries. The confinement term shows a $1/r^2$ dependence of optical energy gap. Fig 4 shows the plot between band gap *vs* grain size (estimated through XRD). $m_h^*$ and $E_g^{bulk}$ can be evaluated using the slope and intercept of the best fit line in Fig 4. In this plot, we have inserted data from previously published data of SnS deposited by thermal evaporation on glass and ITO substrates [33]. Interestingly, all the data points fall on the same trend line, implying (against common impression) that the optical properties of SnS are independent of the substrate used and rather depends on the fabrication conditions only. The estimated $m_h^*$ and $E_g^{bulk}$ are 0.47 $m_o$ and 1.53 eV respectively. This would suggest that the films with grain size greater than 27 nm would behave like bulk SnS sample.

Furthermore, the value of effective mass of the holes lies between 0.2m$_e$ and 0.5m$_e$ as reported by Vidal et. al. [34]. The 0.2m$_e$ effective mass is in direction parallel to the SnS layers and 0.5m$_e$ effective mass charge carriers are moving along the Van der Waals force direction, *i.e.* perpendicular to the layers of SnS. Our value of effective mass would imply that our samples are not perfectly oriented but has a preferential orientation with layers of SnS parallel to the substrate and film surface. This preferred orientation is also evident from our XRD results shown in Fig. 1(a) where the (040) peak is most intense, which indicates that maximum number of crystallites have grown orientation along (040) direction. BAEK, In-Hwan, et al. [35] reported the similar preferred orientation for SnS thin film. Thus, the results that shall be discussed in the subsequent section would list the optical properties of SnS thin films along the Van der Waals force directions i.e., perpendicular to the layers of SnS.

The optical absorption coefficients of the SnS thin films show an exponential tail, known as the Urbach tail, beyond the absorption edge. This tail is related to the number of defects present in the sample and is responsible in giving rise to localized states that extend the gap between conduction band and valance band [36]. The Urbach tail of SnS thin films are calculated using a plot between ln(α) and the energy (hv) of the incident photon and by fitting the following expression [37]:

$$\alpha = \alpha_o exp\left(\frac{hv}{E_U}\right) \qquad (3)$$

Where, α$_o$ is constant and E$_U$ is Urhach tail. Taking the logarithmic value of the both sides of above equation, thus, it can be written an equation of a straight line:

$$ln(\alpha) = ln(\alpha_o) + \left(\frac{hv}{E_U}\right) \qquad (4)$$

Fig.5 shows the plot between ln(α) *vs.* energy (hv) for all samples and the values of $E_U$ are calculated from the reciprocal of the gradient of the line. The estimated values of $E_U$ lie in range of 0.296 to 0.350 eV for all samples are reported in Table 1. M. Ben Mbarek et. al. [38] reported values of E$_U$ around 0.19 eV for SnS films. In our samples, as the film thickness increases, the values of E$_U$ increases. This is expected since, thicker films take longer deposition time which increases defects in the films. A small contribution is decrease in Eg with increase in film thickness due to increased defects can also be appreciated [39].

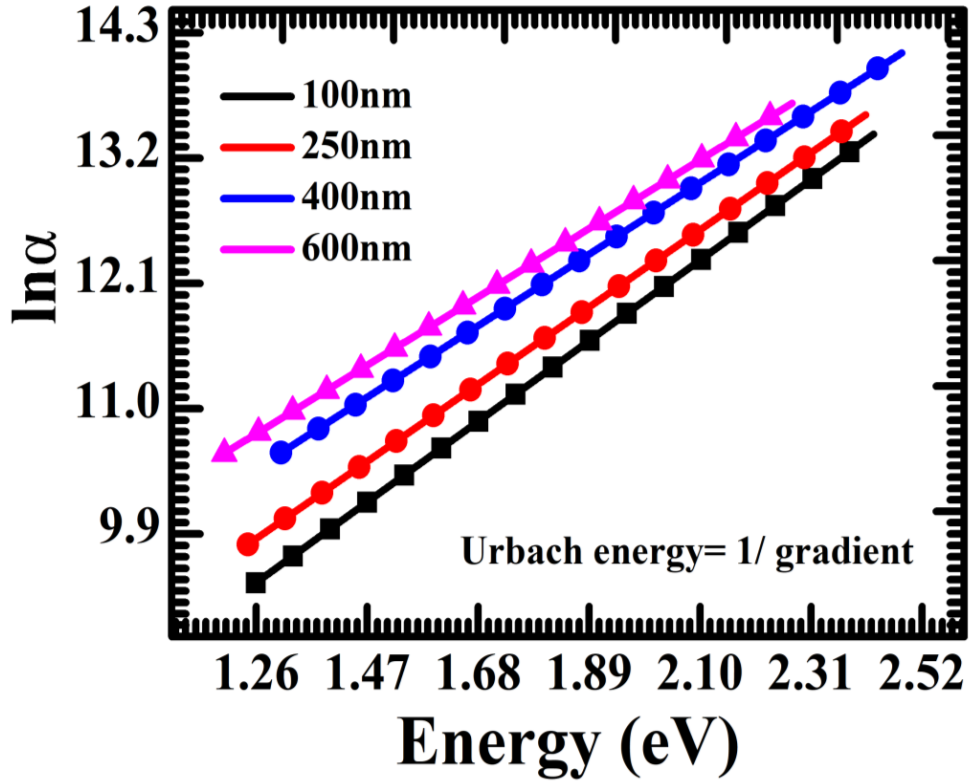

Fig.5 Urbach energy of SnS thin film with thickness of 100 to 600 nm.

Table 1: Optical band gap ($E_g$), Urbach energy ($E_U$), refractive index (n) of SnS films.

| Thickness (nm) | $E_g$ (eV) | $E_U$ (eV) | n (Swanepoel) at 1770 nm |
|---|---|---|---|
| 100 | 2.07 | 0.296 | 1.56 |
| 250 | 1.71 | 0.310 | 1.68 |
| 400 | 1.59 | 0.335 | 1.70 |
| 600 | 1.30 | 0.350 | 1.81 |

### 3.2.3 Refractive Index of SnS Thin Films

Refractive Index of deposited SnS thin film in the spectral region of transparent and weak absorption can be obtained using the Swanepoel's method. Generally, this method is used for analyzing a first approximate value of the refractive index (*n*) of the thin films, in the transparent region according to following expression [21]:

$$n = \sqrt{N_1 + \sqrt{N_1^2 + s^2}} \qquad (5)$$

$$N_1 = \frac{T_M - T_m}{T_M T_m} + \frac{s^2 + 1}{2} \tag{6}$$

where `n' is refractive index, $T_M$ and $T_m$ are the maximum and minimum transmission and 's' is the refractive index of the FTO glass substrate. Fig.6 shows the variation of refractive index with wavelength of SnS thin films with thickness from 100 to 600 nm. The estimated refractive index using Swanepoel's method of SnS films with thickness of 100 to 600 nm are summarized in Table 1. M. S. Selim et. al. [40] reported similar range of refractive index for thickness dependent SnS films. Here as thickness of thin films increases the refractive index increases. In next section, we will try to investigate the possible cause of this variation.

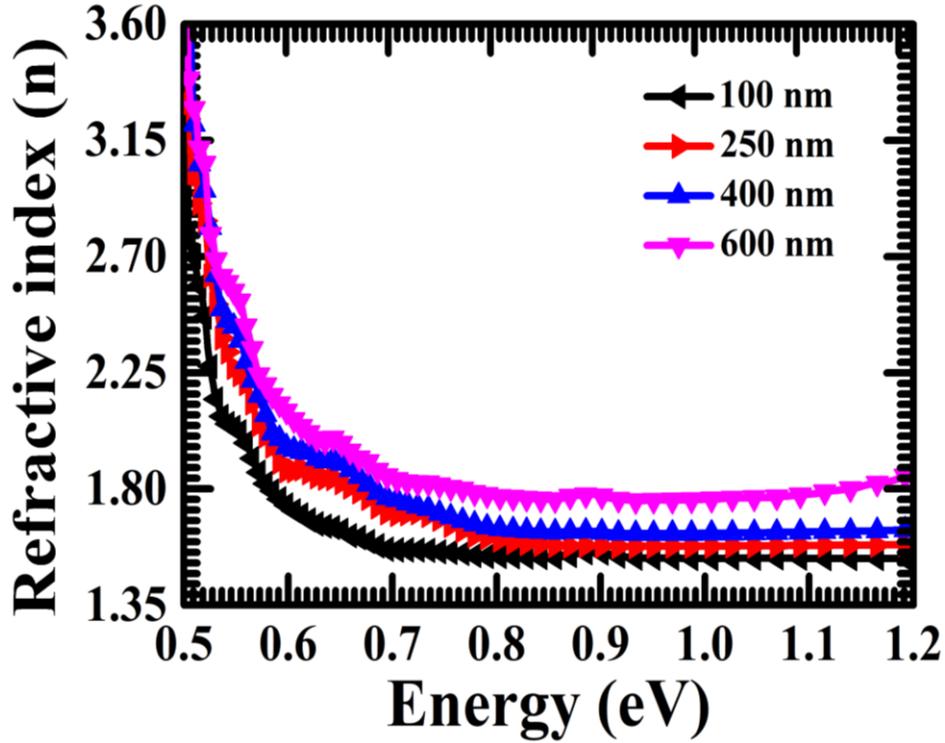

Fig.6 Refractive index of SnS thin film with thickness of 100 to 600 nm.

### 3.2.4 Dispersion Analysis of SnS Thin Films

The Wemple-DiDomenico (WDD) model is used to explain the variation of refractive index near the absorption edge and is successful in giving an insight about the material property [41,42]. The model relates energy dependence of refractive index through the relationship [43]:

$$n^2 = 1 + \frac{E_d E_0}{(E_0^2 - (h\nu)^2)} \tag{7}$$

The value of $E_d$ and $E_o$ can directly determine by plotting $(n^2-1)^{-1}$ against $(h\nu)^2$, as shown in Fig 7(a), which usually gives a linear region. From the intercept ($E_0/E_d$), on the vertical axis and the slope ($1/E_0 E_d$) of this linear part, the dispersion parameters $E_o$ and $E_d$ can easily be calculated. The dispersion parameter $E_d$ is the oscillator strength, which is a measure of the strength of inter-band optical transitions. The oscillator energy $E_o$ is an average optical energy gap. Furthermore, an approximate value of the optical band gap ($E_g$) of SnS thin films might be obtained from the WDD model by using relation $E_g \approx E_0/2$. The estimated value of $E_d$, $E_g$ and $E_0$ using above equation, are summarized in Table 2 for thickness from 100 to 600 nm thick SnS thin films, respectively. The values of $E_g$ such obtained were found to be approximately equal to that estimated by our Tauc's calculation using the absorption data, with values converging for higher thicknesses. The increase in values of $E_d$ is attributed to the increase in coordination number of the atoms. Interestingly, we have combined our data with an earlier work of one of the authors on SnS [36] and find that, $E_d$ decreases with increasing $E_U$, the variation of $E_d$ with $E_U$ is shown in Fig.7(b).] That is, the inter-band transition (probability) strength decreases with increasing Urbach tail. This suggests, along with the major contribution of grain size, increasing defects in SnS with increasing thickness of films also contribute to a red shift in band gap. Also, structural information can be appreciated from $E_o$ and $E_d$'s relation to the moments of the spectrum [44]:

$$E_o^2 = \frac{M_{-1}}{M_{-3}} \quad \& \quad E_d^2 = \frac{M_{-1}^3}{M_{-3}} \tag{8}$$

Here, $M_{-1}$ and $M_{-3}$ are the first and the third negative moments of absorption spectra, which represent the measure of the average bond strength. The two moments $M_{-1}$ and $M_{-3}$ were calculated using the values of dispersion energy parameters are summarized in Table 2. It can be noticed that the values of $M_{-1}$ and $M_{-3}$ increases with increasing film thickness. Together with the Urbach tail's interpretation, we can conclude the comprehensive stress relaxes with decreasing defects, resulting from increasing film thickness. Thus, distance between neighboring atoms increases with increasing film thickness, resulting in an increase in $E_d$ value.

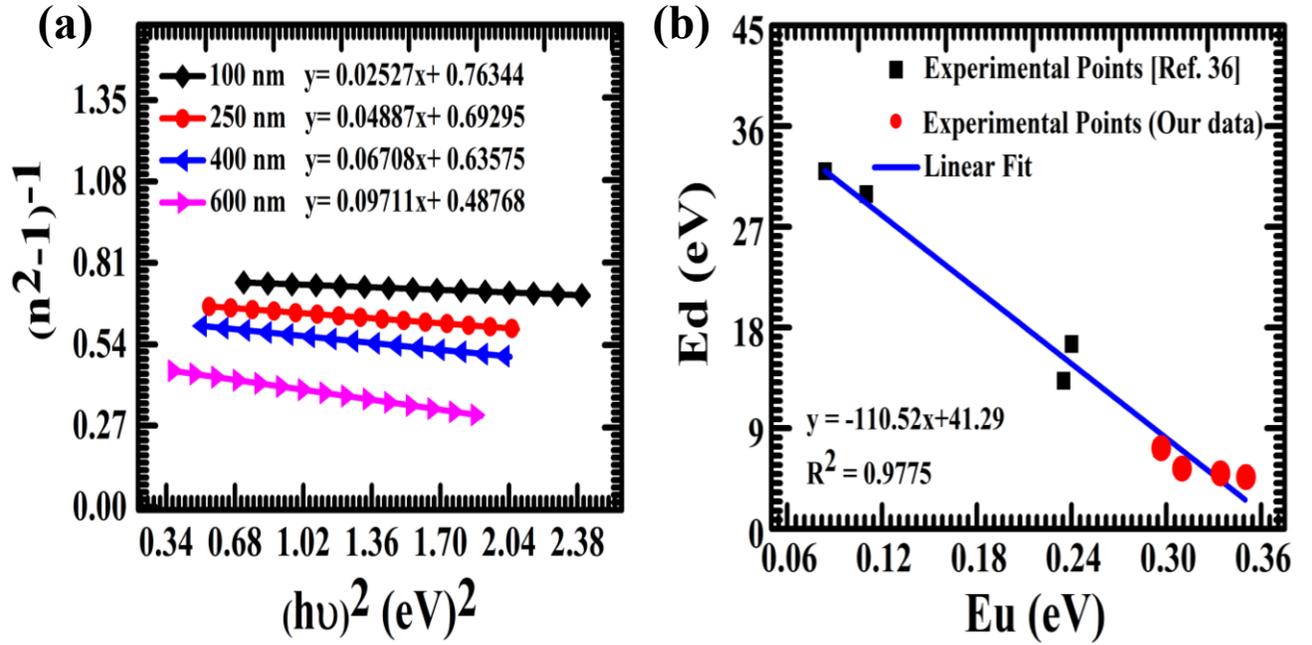

Fig.7(a) Plot of $(n^2-1)^{-1}$ with $(h\nu)^2$ of SnS thin film with thickness of 100 nm to 600 nm, (b) variation of Urbach energy with dispersion energy of SnS thin films.

Lattice energy ($E_l$) is another interesting parameter, which gives information about the strength of the bonds in an ionic compound. Inserting this term in eq. (8), the additional potential that acts on the electrons, restoring it to its mean position when incident light's electric field sends it into oscillation, is taken into account. $E_l$ can be calculated using the following relation [45]:

$$n^2 = 1 + \frac{E_d E_0}{E_0^2 - (h\nu)^2} - \frac{E_l}{(h\nu)^2} \tag{9}$$

At longer wavelength, where $E_o^2 \gg (h\nu)^2$, equation (9) can be written as:

$$n^2 = 1 + \frac{E_d}{E_0} - \frac{E_l}{(h\nu)^2} \tag{10}$$

where $E_o$ and $E_d$ are single-oscillator constants. The parameters $E_l$ is calculated from the slope of plot between $(n^2-1)$ vs. $1/(h\nu)^2$, as shown in Fig.8. The calculated values of the $E_l$ using above equation are summarized in Table 2 for thickness from 100 to 600 nm thick SnS thin films, respectively. The increase in lattice energy with increasing thickness is a measure of increase in difficulty to set phonons into vibration. Thus, explaining the difficulty in electron-phonon interaction and in turn increment in refractive index with films thickness. The increase in refractive index with increasing films thickness may also be due to the improvement of the crystallization or minimization of defects as understood from variation in Urbach tail with increasing film thickness.

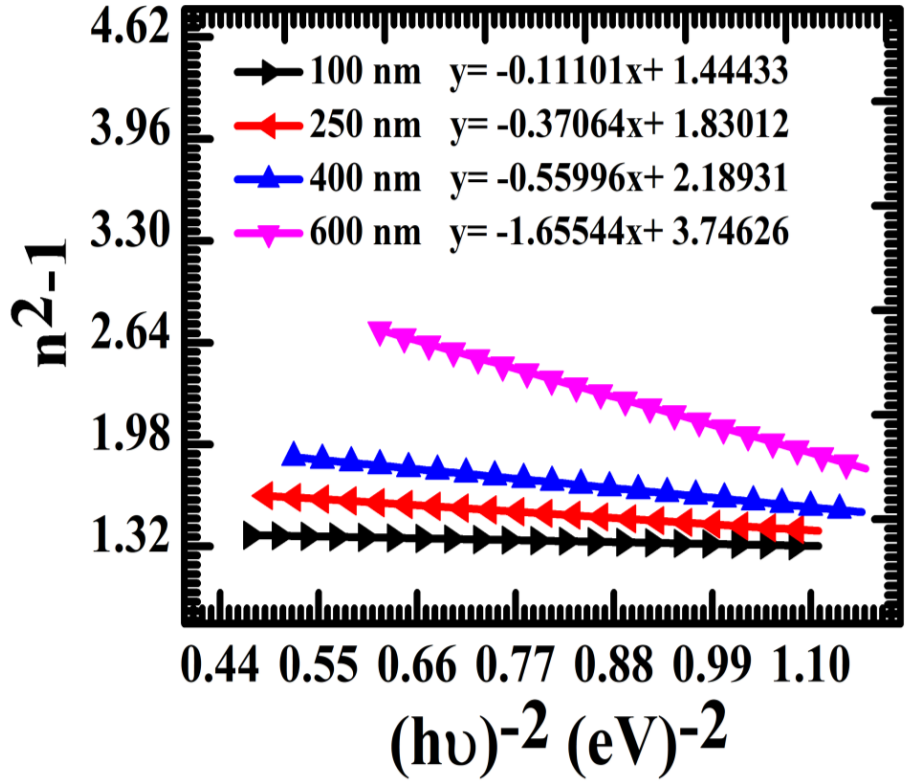

Fig.8 Plot of (n² -1) with $(h\nu)^{-2}$ of SnS thin film with thickness of 100 to 600 nm.

### 3.2.5 High Frequency Dielectric Constant of SnS Thin Films

Eq. (11) relates the real part of the dielectric constant (ε') with wavelength (λ) [46,47]

$$\varepsilon' = n^2 - k^2 = \varepsilon_\infty - \left(\frac{e^2}{4\pi c^2 \varepsilon_0}\right)\left(\frac{N}{m^*}\right)\lambda^2 \quad (11)$$

Where, $c$ is the speed of the light, $N$ the carrier concentration, $m^*$ the effective mass of the charge carriers and $k$ is the extinction coefficient. The extinction coefficient required for the above is calculated using the formula [41]:

$$k = \frac{\alpha\lambda}{4\pi} = \frac{2.303\lambda}{4\pi d}\log(A) \quad (12)$$

Here, $\alpha$ is optical absorption coefficient and $d$ is the film thickness. Fig. 9 shows the plot of $n^2 - k^2$ versus $\lambda^2$ and the linear fit to the data. The relation eq. (11) and (12), allows for the determination of the carrier concentration (*N*) in our SnS samples. We have determined $\varepsilon_\infty$ and *N/m*\* using the information of the intercept and slope of this line. Using the evaluated *m*\* discussed above, the value of *N* was calculated for all SnS thickness and is listed in Table 2. The order of N ($10^{19}$ cm⁻³) is in agreement with reported literature [48]. As the thickness of SnS films increases, the order of carrier concentration remains fairly constant. Data clearly shows that the velocity of light in SnS thin films decreases with increasing thickness (as reflected by increasing `n' or the real part of the dielectric constant, $\varepsilon$'). This is not due to carrier concentration or change in optical density (via $m^*$) but may results from increasing lattice energy and optimization of bond length on account of decreased defects. This would enhance electron-phonon interaction, explaining change in velocity/ wavelength as light travels through SnS.

Surface Plasmons find application in various fields like food safety, lab medicine, environmental monitoring (in all bio-sensing) and in fiber optics etc. While accepted standards of observing them are embedding metal nanoparticles in insulators, recent studies have shown Surface Plasmon Polaritons (SPP) can be found in semiconductors too [49]. The SPP frequency in semiconductors is related to the plasmonic frequency through the relation [50]:

$$\omega_{SPP} = \omega_P \sqrt{\frac{\epsilon_\infty}{\epsilon_\infty + \epsilon_o}} \qquad (13)$$

We can calculate the plasma frequency ($\omega_p$) based on idea at what frequency do the sample's charge carriers resonate with the external oscillating electric field [46]:

$$\omega_p = \sqrt{\frac{N}{m^*}\left(\frac{e^2}{\varepsilon_0}\right)} \qquad (14)$$

Considering that we have determined $\epsilon_\infty \gg \epsilon_o$, we can conclude $\omega_{SPP} = \omega_P$. The estimated value of plasma frequency using eq.14 are summarized in Table 2. The value of plasma frequency ($\omega_p$) is found to be in THz ($10^{14}$ Hz), which is encouraging since it lays in the visible spectra and hence can find application in fiber telecommunication etc.

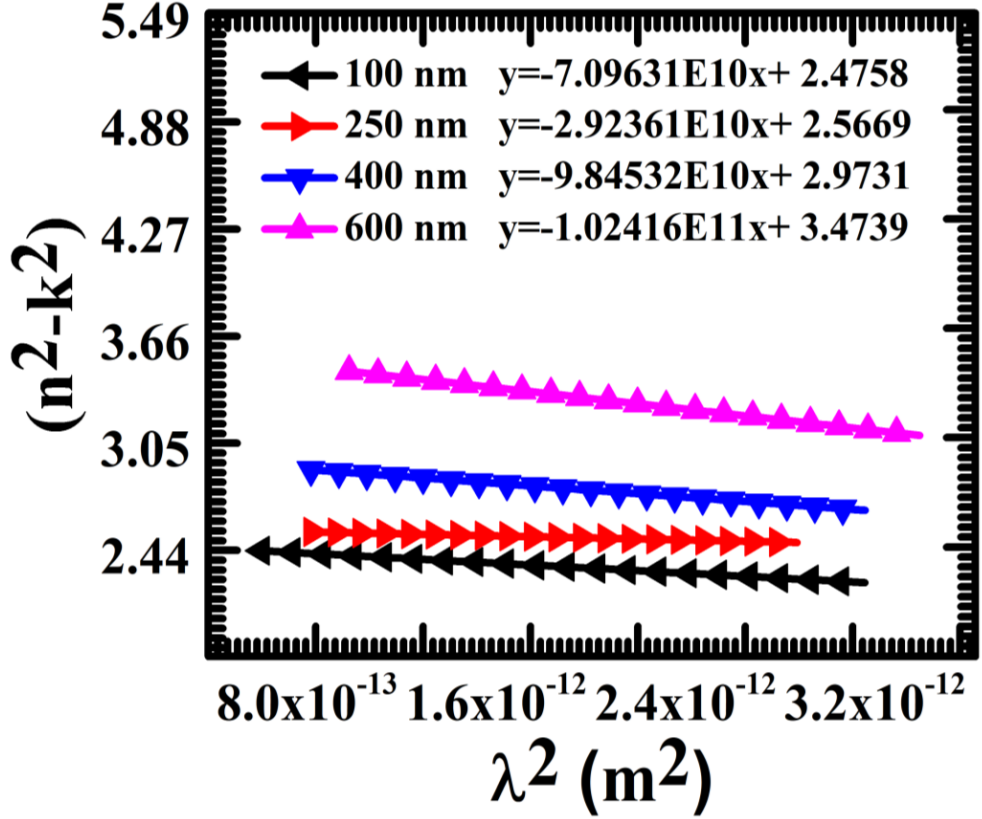

Fig.9: Plot of n²-k² with $\lambda^2$ of SnS thin film with thickness of 100 to 600 nm.

### 3.2.6 Electronic Polarizability of SnS Thin Films

One of the most significant characteristics in examining the optical properties of SnS thin films is the average electronic polarizability of ions. The electronic polarizability ($\alpha_p$) can be derived according to Clausius– Mossotti's equation of dielectric constant, given as [51]:

$$\frac{n^2-1}{n^2+2} = \frac{4N\rho}{3M}\alpha_p \tag{15}$$

Where, $N$ is the Avogadro's number, $\rho$ is the density of material, and $M$ is molecular weight. The photon energy dependence of $(n^2-1)/(n^2+2)$ is shown in Fig.10. Electronic polarizability ($\alpha_p$) values of SnS thin films were estimated by extrapolation of the linear part towards the *y*-axis in the above equation. The estimated electronic polarizability using the above equation are summarized in Table 2 for thickness from 100 to 600 nm thick SnS thin films, respectively. While, dielectric constant is a microscopic property of the material, polarizability is the microscopic or atomic level property. It talks of the dipole induced on application of electric field. The values suggest electrons of SnS are easily set into oscillation as light is incident on the material, however, the electron-phonon interaction reduces velocity of the propagating light in the medium due to strong electron-phonon interaction.

We observed that the evaluated value of polarizability for SnS is appreciable. Hence, we expect the material to show non-linear optical properties. For this we investigate the mathematically evaluated values of $\chi^{(3)}$ and $n_2$ in the next section.

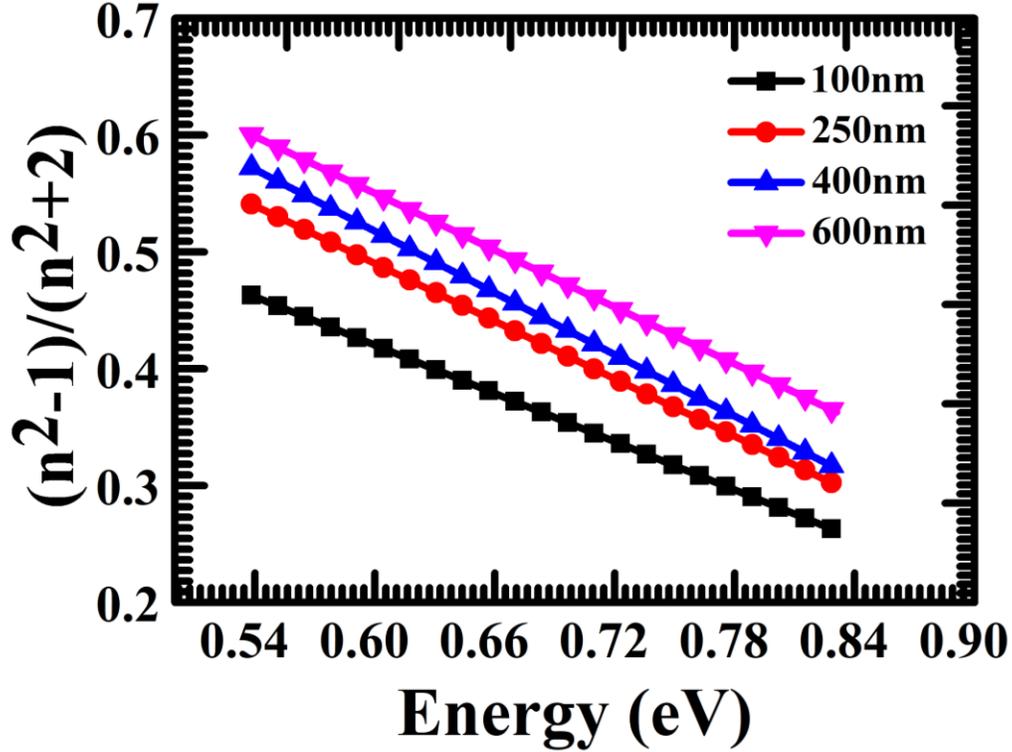

Fig.10: Plot of $(n^2-1)/(n^2+2)$ vs energy of SnS thin film with thickness of 100 to 600 nm.

### 3.2.7 Optical Nonlinearity in SnS Thin Films

The third order of nonlinear optical susceptibility ($\chi^{(3)}$) is an important optical parameter, which provides the information about the strength of the chemical bonds between the molecules of deposited SnS thin films. Tichy and Ticha et. al [52] has reported empirical relation between nonlinear optical susceptibility and nonlinear refractive index ($n_2$) using generalized Miller´s rule and linear refractive index (n). $\chi^{(3)}$ and $n_2$ of SnS thin films with different thickness can be calculated using the following empirical relation[53]:

$$\chi^{(3)} = A \left(\frac{n^2-1}{4\pi}\right)^4 \tag{16}$$

Where, A= $1.7 \times 10^{-10}$ for $\chi^{(3)}$ measured in esu, the $n_2$ is related to the $\chi^{(3)}$ by

$$n_2 = \frac{12\pi \chi^{(3)}}{n_0} \tag{17}$$

Estimated values of $\chi^{(3)}$ were found in the range of 0.288-1.833× $10^{-13}$esu. similar to that reported by P Patra et. al. [54] for CdS thin film, I.M.El Radaf et. al. [41] report the similar value for $Cu_2MnGeS_4$ thin flms. The values of $\chi^{(3)}$ and $n_2$ increases with increasing thickness of the films similar to that seen in $E_d$ and $E_l$. The variation of nonlinear optical susceptibility and nonlinear refractive index with material properties has been a subject of study for the last couple of decades. It has been established that $n_2$ and the $Re|\chi^{(3)}|$ are both related and is proportional to $E_g^{-4}$[55]. The linear variation of nonlinear optical susceptibility and nonlinear refractive index with $E_g^{-4}$ of SnS thin film is shown in Fig. 11. The high non-linearity suggests thick films of SnS has potential in application of fiber telecommunication or where harmonic generation is required.

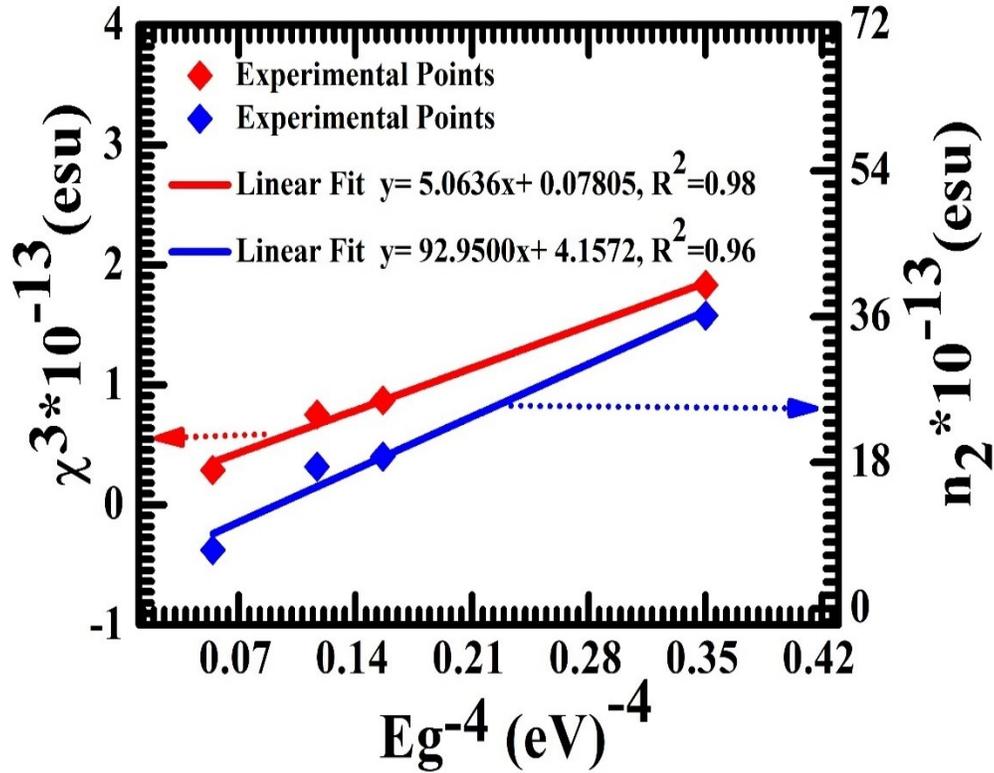

Fig.11: Variation of nonlinear optical susceptibility and nonlinear refractive index with $E_g^{-4}$ of SnS thin film.

Table 2: Optical dielectric and dispersion analysis of SnS thin films.

| Film Parameter | Thickness (nm) | | | |
|---|---|---|---|---|
| | 100 | 250 | 400 | 600 |
| $E_d$ (eV) | 7.20 | 5.43 | 4.87 | 4.59 |
| $E_0$ (eV) | 5.49 | 3.76 | 3.07 | 2.24 |
| $E_g$ (eV) WDDA | 2.74 | 1.88 | 1.53 | 1.12 |
| $M_{-1}$ | 1.311 | 1.444 | 1.586 | 2.049 |
| $M_{-3}$ (eV)$^2$ | 0.043 | 0.102 | 0.168 | 0.408 |
| $E_l$ (eV) | 0.111 | 0.370 | 0.559 | 1.655 |
| N (/cm$^3$) | 11.91×10$^{18}$ | 4.91×10$^{18}$ | 16.51×10$^{18}$ | 5.53×10$^{18}$ |
| $\varepsilon_\infty$ | 2.47 | 2.56 | 2.97 | 3.47 |
| $\omega_p$ (Hz) | 2.81×10$^{14}$ | 1.78×10$^{14}$ | 3.3×10$^{14}$ | 0.614×10$^{14}$ |
| $\alpha_p$ ×10$^{-23}$ | 2.99 | 3.53 | 3.76 | 3.81 |
| $\chi^{(3)}$ ×10$^{-13}$ | 0.288 | 0.753 | 0.871 | 1.833 |
| $n_2$ ×10$^{-13}$ | 7.23 | 17.51 | 18.75 | 36.16 |

## 4. Conclusions:

The SnS thin films have been deposited on FTO substrates through thermal evaporation technique at room temperature to analyze the influence of film thickness on the optical behavior of SnS and obtained results are corelated with SnS films grown on ITO and glass substrates. Our study suggests that the properties of SnS is independent of the substrate-type. Also, irrespective of the substrate used, thermally evaporated SnS films tend to orient itself such that the layers are parallel to the substrate with the Van der Waals forces acting perpendicular to the substrate. Moreover, it is found that the SnS film with grain size greater than 27 nm is found to behave as in bulk state. Refractive index and extinction coefficient increased with an increase of thickness and the value of refractive index is in the range 1.56 to 1.81 at 1770 nm. In addition, the dielectric constants, relaxation time and the single oscillator energy and the dispersion energy are estimated using Wemple–DiDomenico approach. The single oscillator energy was found in the range of 5.49–2.24 eV. The dispersion energies of the thin films were in the range of 7.20–4.59 eV. The nonlinear optical susceptibility and nonlinear refractive index ($n_2$) were calculated by using generalized Miller´s rule and linear refractive index (n). These nonlinear parameter values were found to increase with increasing thickness. All the results point to an ease with which electrons in SnS are sent into oscillation as light falls on preferred orientation, however, the electron-phonon interaction reduces the velocity of the propagating light in the SnS media.

## Acknowledgment

BKS sincerely acknowledges financial support from the institution of Eminence (IoE) BHU grant number-6031, Govt. of India. Vinita acknowledges financial support as Junior Research Fellowship (JRF) from University Grant Commission (UGC) India.